\documentclass[reprint,aps,pra,a4,10pt]{revtex4-1}
\usepackage{amsmath} 
\usepackage{color, soul}
\usepackage{graphicx}
\usepackage{amssymb}
\usepackage{wrapfig}
\usepackage{hyperref,url}

\newcommand{\be}{\begin{equation}} 
\newcommand{\ee}{\end{equation}}
\newcommand{\bea}{\begin{eqnarray}} 
\newcommand{\eea}{\end{eqnarray}}
\newcommand{\bc}{\begin{center}}
\newcommand{\ec}{\end{center}}

\begin{document}
\title{Algebraic theory of endohedrally confined diatomic molecules: application to H$_2$@C$_{60}$}
\author{Lorenzo Fortunato}
\affiliation{Dipartimento di Fisica e Astronomia ``G.\ Galilei'', Universit\`a di Padova\\
and I.N.F.N.- Sez. di Padova; v. Marzolo, 8, I-35131, Padova, ITALY}

\author{Francisco P\'erez-Bernal}
\affiliation{Grupo de investigaci\'on en F\'{\i}sica Molecular, At\'omica y Nuclear (GIFMAN-UHU), Unidad Asociada al CSIC. Depto.\ de Ciencias Integradas, Universidad de Huelva, 21071 Huelva, SPAIN} 

\date{\today}

\begin{abstract}
  A simple and yet powerful approach for modeling the structure of
  endohedrally confined diatomic molecules is introduced. The theory,
  based on a $u(4)\oplus u(3)$ dynamical algebra, combines $u(4)$, the vibron
  model dynamical algebra, with a $u(3)$ dynamical algebra that models
  a spherically symmetric three dimensional potential. The first
  algebra encompasses the internal roto-vibrations degrees of freedom
  of the molecule, while the second takes into account the confined
  molecule center-of-mass degrees of freedom. A resulting subalgebra
  chain is connected to the underlying physics and the model is
  applied to the prototypical case of H$_2$ caged in a fullerene
  molecule. The spectrum of the supramolecular complex H$_2$@C$_{60}$
  is described with a few parameters and predictions for not yet
  detected levels are made.  Our fits suggest that the quantum numbers
  of a few lines should be reassigned to obtain better agreement with
  data.
\end{abstract}


\maketitle

\section{Introduction}
Supramolecular species in which a guest atom or molecule is inserted
in the interior of a host molecule (usually fullerenes), are known as
endohedral compounds, and form systems that are bound by the pure
confinement rather than by intra-molecular forces. The first
endohedral compounds obtained consisted of trapped metal atoms
\cite{Chai:1991} followed by endofullerenes with a trapped molecule \cite{Ito:2008}. These systems display a full gamut of
quantum effects, because the confinement of the molecule results in the
splitting of the translational degrees of freedom of the incarcerated
molecule center of mass and their coupling with roto-vibrational ones. A fundamental
breakthrough that has allowed the application of different spectroscopic tools to molecular
endofullerenes has been the achievement of high reaction yields in their synthesis
using the so called "molecular surgery" (See e.g.\ Refs.~\cite{Levitt:13,Komatsu:2013} and references therein). 
Komatsu and coworkers have
presented the synthesis of the endohedral species H$_2$@C$_{60}$, that is the subject
of the present work \cite{Komatsu:05}. Another impressive
step forward in this area has been Murata's group achievement, using
similar experimental techniques, of the synthesis of a closed water
endofullerene \cite{Kurotobi:2011} and the recent encapsulation of hydrogen fluoride inside C$_{60}$ \cite{Krachmalnicoff:16}.

Significant experimental and theoretical research efforts have been
devoted to the elucidation of the spectral properties of
H$_2$@C$_{60}$ due to the remarkable quantum
effects that link roto-vibrational and translational degrees of
freedom, coming into play once the diatomic molecule is trapped
into the buckyball. In the case of incarcerated H$_2$, the well known
existence of two allotropes of the hydrogen molecule,
\textit{para}-H$_2$ and \textit{ortho}-H$_2$, make this compound a
valuable tool for explorations in spin chemistry
\cite{Turro:10}. These fascinating characteristics of the
supramolecular complex H$_2$@C$_{60}$ have stimulated remarkable
experimental efforts with different techniques
\cite{Mamone:11,Levitt:13}, mainly Nuclear Magnetic Resonance (NMR)
\cite{Komatsu:05,Carravetta:07, Turro:10}, InfraRed (IR)
\cite{Mamone:09,Ge:11, Room:13}, and Inelastic Neutron Scattering
(INS) \cite{Horsewill:10,Horsewill:12,Horsewill:13,Xu:14,
  Mamone:16}. In particular, an INS spectroscopy selection rule of
H$_2$@C$_{60}$ has been recently discovered
\cite{Xu:14,Xu:15,Poirier:15}. The search for an adequate description
of the structure and the peculiar properties of this endohedral
species has provoked intense theoretical efforts
\cite{Xu:08,Xu:08bis,Xu:11bis, Xu:11, Xu:15, Poirier:15}. This system
represents an almost ideal testing ground for theories because it
couples the simplest diatomic molecule with an almost perfect
spherical cage (the icosahedral symmetry can be neglected for most
practical purposes). The neutral molecule retains its bound character
but, at the same time, it is affected by the presence of the fullerene: its
motion is confined and quantized due to the interaction with the
cage, a situation that can be fully explored by powerful and simple
symmetry-guided models.

Measurements of the IR spectrum of H$_2$@C$_{60}$ from low
temperatures up to room temperatures have been performed
\cite{Mamone:09,Ge:11,Room:13}. Combining IR spectroscopy data
and INS results, the lowest portion of the endohedral compound
spectrum has been measured with sufficient detail to allow the
experimental underpinning of the differences, shifts and splitting of
the levels with respect to the free H$_2$ counterpart. The spectrum of
the confined H$_2$ molecule has been interpreted in terms of a very
accurate, though computationally involved, five dimensional
phenomenological model \cite{Xu:08, Xu:08bis, Xu:11bis, Xu:11}. While these five dimensional
calculations are accurate and can be used to conveniently
describe the observations and to make guesses about still unobserved
excited states, it is not completely obvious what is the origin of the
perturbations in the potential energy terms. For example in
\cite{Xu:08bis} the authors use Lennard-Jones potentials for each H-C
pair in the complex, realizing that the use of an angular momentum
quantum number associated with a harmonic motion of the molecule
inside the cage is indeed appropriate. This fact supports the
convenience of a computationally inexpensive symmetry-based approach like the one we suggest.

We will describe our algebraic approach in Sect.~II, discuss the methodology and the 
fits to a set of experimental lines in Sect.~III, and draw conclusions in Sect.~IV.

\section{Algebraic Approach}
Stimulated by the success of the existing approach \cite{Mamone:09, Ge:11}, 
with the aim of
obtaining a simple model that encompasses the main physical
ingredients for such an enticing system, we propose in the present
manuscript an algebraic theory for the quantum modes of a diatomic
molecule confined in an isotropic three-dimensional cage. Symmetry
considerations constitute the guiding principle that inspires the
treatment of the energy terms obtained from a Hamiltonian operator
that includes molecular roto-vibrational and center-of-mass modes, and
the coupling of these two subsystems.  The rotations and vibrations of
the diatomic molecule are described within the vibron model
\cite{Iachello:81, Iachello:94, Frank:05}, that amounts to a $u(4)$
Lie algebra arising from the bilinear products of scalar $s,s^\dag$
($\ell=0$) and vector $p_\mu,p_\mu^\dag$ ($\ell=1, \mu = \pm 1, 0$)
boson operators \cite{Iachello:81}. The fullerene cage is modeled as
a spherical three dimensional well and can be dealt with a
$u(3)$ Lie algebra, arising from a vector boson operator
$q_\mu,q_\mu^\dag$ ($\ell=1, \mu=\pm 1, 0$) \cite{Iachello:06}. Taking
this into consideration we invoke an algebraic model based on the
direct sum Lie algebra $u_p(4) \oplus u_q(3)$ to describe the
intrinsic modes of excitations of the supramolecular complex
H$_2$@C$_{60}$, where we use the subindexes $p$ and $q$ to distinguish
the two different sets of degrees of freedom. Our symmetry-inspired
scheme should be desirable for at least the following peculiar
features: (i) it gives a simple framework that singles out what are
the linearly independent energy terms and their connection with
physical operators; (ii) it gives a natural explanation for the
interaction between translational and roto-vibrational degrees of
freedom responsible for term energy splittings; (iii) it also gives a
natural explanation for the specific radial and angular dependence
($\{R,\Omega, \Omega_s\}$ in the notation of Refs.\ \cite{Mamone:09,
  Ge:11}) of the terms that have been found to contribute to the
expansion of the coupling potential function; (iv) it treats on an
equal footing \emph{para}- and \emph{ortho}-H$_2$ states; (v) there is no need to
find separate sets of parameters for each vibrational band, a single
fit encompasses all vibrational states simultaneously; and (vi) it is
computationally inexpensive: the matrix elements of each operator are
known in closed form and the diagonalization can be performed exactly
and rapidly. In addition,
it yields precise predictions for higher lying modes that, although
unseen heretofore, might be investigated in future.

A model that shares a similar algebraic structure, with a dynamical
algebra $u(7) \supset u(3) \oplus u(4)$, has been used in the description
of hadronic structure in terms of quark building blocks
\cite{Bijker:94,Iachello:06}. In that model the $u(7)$ algebra arises from two
Jacobi coordinate vectors that describe quarks inside a baryon plus a
scalar boson and it is used for the spatial part of the description
that must be supplemented by a fermionic part containing the
flavor, spin and color degrees of freedom. 
While the algebraic
structure is very similar, clearly the physics behind the model is
completely different. 

Our model provides a complete mathematical characterization of all
possible interactions that comply with the underlying symmetries and
therefore naturally gives a hint of the various
physical mechanisms that might generate them. We will confine the
present discussion to identifying the most important terms and return
to the laborious task of a complete classification in a longer paper \cite{Perez:16tobe}.

Among the many possible subalgebra chains, we consider the following dynamical symmetry
\begin{widetext}
\begin{equation}
\begin{array}{cccccc}
 u_p(4) \oplus u_q(3) & \supset  &so_p(4) \oplus u_q(3)  \supset& so_p(3) \oplus so_q(3)  & \supset so_{pq}(3)& \supset so_{pq}(2)   \\
 ~N_p    \hfill N_q~  &          &~\omega\hfill ~              &   ~J     \hfill L~      &       \Lambda    &         M_\Lambda  
\end{array}~,
\label{basis}
\end{equation}
\end{widetext}
\noindent where we have used the $so(4)$ limit of the vibron
model \cite{Iachello:81,Iachello:94,Frank:05,Iachello:06} and where the
second line gives the quantum numbers associated with the Casimir
operators of each algebra. With the {\it proviso} that $\omega$ is
related to the vibrational quantum number $v$ through $v={\textstyle
  \frac{1}{2}}(N_p-\omega)$, the set $(vJN_qL\Lambda)$ corresponds to
the quantum numbers used so far in theoretical investigations. The
basis states can therefore be labeled, very similarly to Refs.\
\cite{Xu:08,Xu:08bis, Ge:11, Xu:13, Horsewill:10, Horsewill:12,
  Mamone:09, Mamone:11}, as $|N_p v J; N_q L;  \Lambda \rangle$.

 The quantum numbers follow
the well-known branching rules \cite{Frank:05,Iachello:06}
\begin{align}
\omega &= N_p, N_p-2, \ldots, 1 \text{ or } 0~,\nonumber\\
J &= 0, 1, \ldots, \omega~,\nonumber\\
L &= N_q, N_q -2,\ldots, 1 \text{ or } 0~,\\
\Lambda &= |J-L|, |J-L| +1, \ldots, J+L~.\nonumber\\
M_\Lambda &= -\Lambda, -\Lambda +1, \ldots, \Lambda-1, \Lambda~.\nonumber
\end{align}

The total Hamiltonian can be written as 
\begin{equation} 
\hat H_{endo} = \hat H_{u_p(4)} + \hat H_{u_q(3)} + \hat H_{Coupl}~,
\label{totH}
\end{equation}
where the first term represents the vibron model Hamiltonian for
rotations and vibrations of a diatomic molecule \cite{Iachello:94}, the second is the quantized motion of the molecular
center-of-mass inside the three-dimensional spherically-symmetric confining potential, and the last term includes molecule-cage couplings.

The $u(4)$ vibron model Hamiltonian can be modeled as 
\begin{equation}
\hat H_{u_p(4)}= \hat H_{so(4)} + \hat H_{Dun}~,
\label{hup4}
\end{equation}
where the first term contains the two-body Casimir operators of the $so(4)$ dynamical symmetry and the second includes two higher-order terms in a Dunham-like expansion \cite{Iachello:94, Frank:05} where the first term represents a centrifugal correction and the second a rotation-vibration coupling.
\begin{align}
\hat H_{so(4)} &= E_0 +\beta\, \hat C_2[so_p(4)] + \gamma\, \hat C_2[so_p(3)]~,\label{so4ham}\\
\hat H_{Dun} &= \gamma_2\hat C_2[so_p(3)]^2+\kappa\,  \hat C_2[so_p(4)]\hat C_2[so_p(3)]~.\label{corrham}
\end{align}
The Casimir operators in Eqs.\ \eqref{so4ham} and \eqref{corrham} are diagonal in the chosen basis \eqref{basis}
\begin{align}
\langle \alpha |\hat C_2[so_p(4)]|\alpha \rangle =& \omega (\omega+2)~,\nonumber\\
\langle \alpha |\hat C_2[so_p(3)]|\alpha \rangle =& J(J+1)~,\label{pmatel}\\
\langle \alpha |\hat C_2[so_p(4)]\hat C_2[so_p(3)]|\alpha \rangle =& \omega (\omega+2)J(J+1)~,\nonumber
\end{align}
\noindent where $|\alpha \rangle = |N_p v J; N_q L;  \Lambda \rangle$.

The energy formula obtained for $\hat H_{u_p(4)}$ is
\begin{align}
E_{u_p(4)}=& E_0 +\beta\, \omega (\omega+2) + \gamma\, J(J+1) \nonumber\\
&+ \gamma_2\Bigl[ J(J+1)\Bigr]^2+\kappa\,  \Bigl[ \omega(\omega+2)J(J+1)\Bigr]~,
\label{enII}
\end{align}
where $\omega= N_p, N_p-2,\ldots, 1 ~{\mbox or }~  0$ or, alternatively, 
$v=0,1,\ldots,{\textstyle \frac{1}{2}} (N_p-1) ~{\mbox or }~ {\textstyle \frac{1}{2}} N_p$
and  $J=0,1,\ldots, \omega $.

 The parameters in Eq.\ \eqref{enII} are free parameters that can be adjusted to optimize the agreement with experimental data and can be put in direct correspondence with those defined in the  approach of Refs.\ \cite{Mamone:09,Ge:11}
 
The center-of-mass degrees of freedom Hamiltonian, within the $u_q(3)$ dynamical symmetry, is 
\begin{equation}
  \hat H_{u_q(3)}= a\, \hat C_1[u_q(3)] +b\, \hat C_2[u_q(3)]+c\, \hat C_2[so_q(3)]~,
  \label{cageham}
\end{equation}
\noindent where the first term is the number of $q$ bosons and would
be the only term in case that the confining potential were an
isotropic 3D harmonic oscillator, the second term is an anharmonic
correction, and the third term is the H$_2$ center-of-mass centrifugal
energy.

Again the Casimir operators are diagonal in the chosen basis 
\begin{align}
\langle \alpha |\hat C_1[u_q(3)]|\alpha \rangle =& N_q~,\nonumber\\
\langle \alpha |\hat C_2[u_q(3)]|\alpha \rangle =& N_q^2~,\label{qmatel}\\
\langle \alpha |\hat C_2[so_q(3)]|\alpha \rangle =& L(L+1)~,\nonumber
\end{align}
where  $|\alpha \rangle = |N_p v J; N_q L;  \Lambda \rangle$. The free
parameters are $a$, $b$, and $c$ and the spectrum associated to the
center-of-mass degrees of freedom can be written down in this approach as
\begin{equation}
  E_{u_q(3)}= a\, N_q+  b\, N_q^2+ c\, L(L+1)~,
  \label{hospec}
\end{equation}
where $N_q$ is the eigenvalue of the number of quanta operator and $L$ is the orbital angular momentum of the whole confined 
particle (\textit{viz}.\ the center of mass of the H$_2$ molecule) inside the fullerene cage.

\subsection{Diatomic Molecule and Spherical Cage Coupling}
The guest diatomic molecule and the cage interact through a number of different physical mechanisms that can be traced back to scalar operators built out of the elements of the different algebras. Even at this level, the model is quite rich, therefore one needs to select the most important operators guided by some physical principle and intuition, rather than looking for global fits that would entail too many parameters.
We have found that the relevant terms imply Quadrupole-Quadrupole couplings.

The algebraic scheme entails two sets of quadrupole operators, namely
$\hat Q_p=[p^\dag\times\tilde p]^{(2)}$, the quadrupole operators of
$u_p(4)$, and $\hat Q_q=[q^\dag\times\tilde q]^{(2)}$, the quadrupole
operators of $u_q(3)$. The former describes the intrinsic (non-null if
$J\ne0$) quadrupole of the H$_2$ molecule, while the latter can be
associated with the quadrupole deformation of the probability
amplitude of the whole molecule inside the spherical cage. A scalar
coupling can be built from these two operators as $[\hat
Q_p^{(2)}\times \hat Q_q^{(2)}]^{(0)}$, which is the basis for the
coupling term in the Hamiltonian \eqref{totH}. In addition, following
the spirit of a Dunham expansion, further terms can be considered that
lead us to select the following coupling Hamiltonian: 
\begin{widetext}
\begin{equation}
\hat H_{Coupl} = \vartheta_{pq} [\hat Q_p^{(2)}\times \hat Q_q^{(2)}]^{(0)} + \vartheta_{pqw} \left[\hat C_2[so_p(4)][\hat Q_p^{(2)}\times \hat Q_q^{(2)}]^{(0)}+[\hat Q_p^{(2)}\times \hat Q_q^{(2)}]^{(0)}\hat C_2[so_p(4)]\right]+ v_{pq} \hat C_1[u_q(3)]\hat C_2[so_p(4)]~.\label{hcoup}
\end{equation}
\end{widetext}

The parameters $\vartheta_{pq}$, $\vartheta_{pqw}$, and $v_{pq}$ can
be used to adjust the interaction strengths. The most important
finding about the $[\hat Q_p^{(2)}\times \hat Q_q^{(2)}]^{(0)}$
quadrupole-quadrupole interaction is that it lifts the degeneracy of
$\Lambda \ne 0$ multiplets, giving the correct and unusual ordering
seen in experiments.  For example, the triplet of states with $J=1$,
$N_q=L=2$ has the ordering $\Lambda=2,3,1$ that cannot be due to a
scalar coupling of the rotational and translational angular momentum. In fact, once J (that is the rotational angular momentum) and L (the translational angular momentum) are set, a term of the form $\vec J \cdot \vec L$ always gives a splitting of the levels $\mid J-L\mid < \Lambda< J+L$ that strictly follows an increasing or decreasing ordering depending on the sign of the strength constant.

Following the appendix of Ref.\ \cite{shalit1963nuclear} or Ref.\ \cite{Edmonds}, the matrix elements of the scalar coupling of the $\hat Q_{p}$ and $\hat Q_{q}$ quadrupole operators are
\begin{widetext}
\begin{equation}
\langle N_p \omega J;N_q  L;  \Lambda|[\hat Q_p^{(2)}\times \hat Q_q^{(2)}]^{(0)}| N_p \omega' J';N'_q L';  \Lambda' \rangle = (-1)^{L+\Lambda+J'} \sqrt{5} \left\{\begin{array}{ccc}J & L & \Lambda\\L'&J'&2\\ \end{array}\right\}\langle N_q L ||\hat Q_q||N'_q L'\rangle\langle\omega J ||\hat Q_p||\omega' J'\rangle\delta_{\Lambda,\Lambda'}~~.
\label{qqelements}
\end{equation}
\end{widetext}
Once we separate the molecular and cage degrees of freedom, the reduced matrix elements of the molecular ($\hat Q_p$)  and center-of-mass ($\hat Q_q$) quadrupole degrees of freedom are \cite{Frank:05}
\begin{widetext}
\begin{align*}
\langle N_q L ||\hat Q_q||N_q L\rangle =& (2N_q + 3)\sqrt{\frac{L(L+1)(2L+1)}{6(2L-1)(2L+3)}}~,\nonumber\\
\langle N_q L+2 ||\hat Q_q||N_q L\rangle =& \sqrt{\frac{(N_q-L)(N_q+L+3)(L+1)(L+2)}{(2L+3)}}~,\\
\langle \omega 0 ||\hat Q_p||\omega 0\rangle =& 0~,\nonumber\\
\langle \omega J ||\hat Q_p||\omega J\rangle =& (N_p + 2)\left(1+\frac{J(J+1)}{\omega(\omega+2)}\right)\sqrt{\frac{J(J+1)(2J+1)}{6(2J-1)(2J+3)}}~,\nonumber\\
\langle \omega J+2 ||\hat Q_p||\omega J\rangle =& (N_p + 2)\sqrt{\frac{(\omega-J-1)_2(\omega+J+2)_2(J+1)(J+2)}{4\omega^2(\omega+2)^2(2J+3)}}~,\\
\langle \omega+2 J-2 ||\hat Q_p||\omega J\rangle =& \sqrt{\frac{(N_p-\omega)(N_p+\omega+4)(\omega-J+1)_4J(J-1)}{16(\omega+1)_3(\omega+2)(2J-1)}}~,\nonumber\\
\langle \omega+2 J ||\hat Q_p||\omega J\rangle =& \sqrt{\frac{(N_p-\omega)(N_p+\omega+4)(\omega-J+1)_2(\omega+J+2)_2J(J+1)(2J+1)}{24(\omega+1)_3(\omega+2)(2J-1)(2J+3)}}~,\nonumber\\
\langle \omega+2 J+2 ||\hat Q_p||\omega J\rangle =& \sqrt{\frac{(N_p-\omega)(N_p+\omega+4)(\omega + J+2)_4(J+1)(J+2)}{16(\omega+1)_3(\omega+2)(2J+3)}}~,\nonumber
\end{align*} 
\end{widetext}
\noindent where we introduce the Pochhammer symbol $(a)_b = a(a+1)\cdots(a+b-1)$.

The matrix elements for the other two operators in the coupling term \eqref{hcoup}, 
 $\left[\hat C_2[so_p(4)][\hat Q_p^{(2)}\times \hat Q_q^{(2)}]^{(0)}+[\hat Q_p^{(2)}\times \hat Q_q^{(2)}]^{(0)}\hat C_2[so_p(4)]\right]$ and $\hat C_1[u_q(3)]\hat C_2[so_p(4)]$ are trivially computed using Eqs.~(\ref{pmatel}), (\ref{qmatel}), and (\ref{qqelements}).

 Another relevant consideration with regard to the
 quadrupole-quadrupole coupling is the following: if one defines a
 total quadrupole operator as the sum of the two effects, $\hat Q_T=
 \hat Q_p+ \hat Q_q$ and takes the ratio of the expectation values of
 this in the first two excited states, namely $|A \rangle = |00111
 \rangle$ and $|B \rangle = |01001 \rangle$, the resulting expression,
$\langle Q\rangle_A / \langle Q\rangle_B = (N_p+2+2/N_p)/3~,$
depends only on $N_p$, the label of the totally symmetric
representation of $u_p(4)$ that sets the available Hilbert space for
the roto-vibrational degrees of freedom, thus giving an alternative
to the usual methods of assessing this parameter \cite{Iachello:94}.

\section{Experimental Data and Fit Results}
We have extracted from the literature a total of 71 line positions,
compiling a database that includes 55 IR transitions \cite{Ge:11} and
16 INS transitions \cite{Horsewill:12}. In these references, lines
have been assigned with initial and final quantum numbers on the basis
of experimental evidence and theoretical models. 

\begin{figure}
\includegraphics[width=0.55\textwidth]{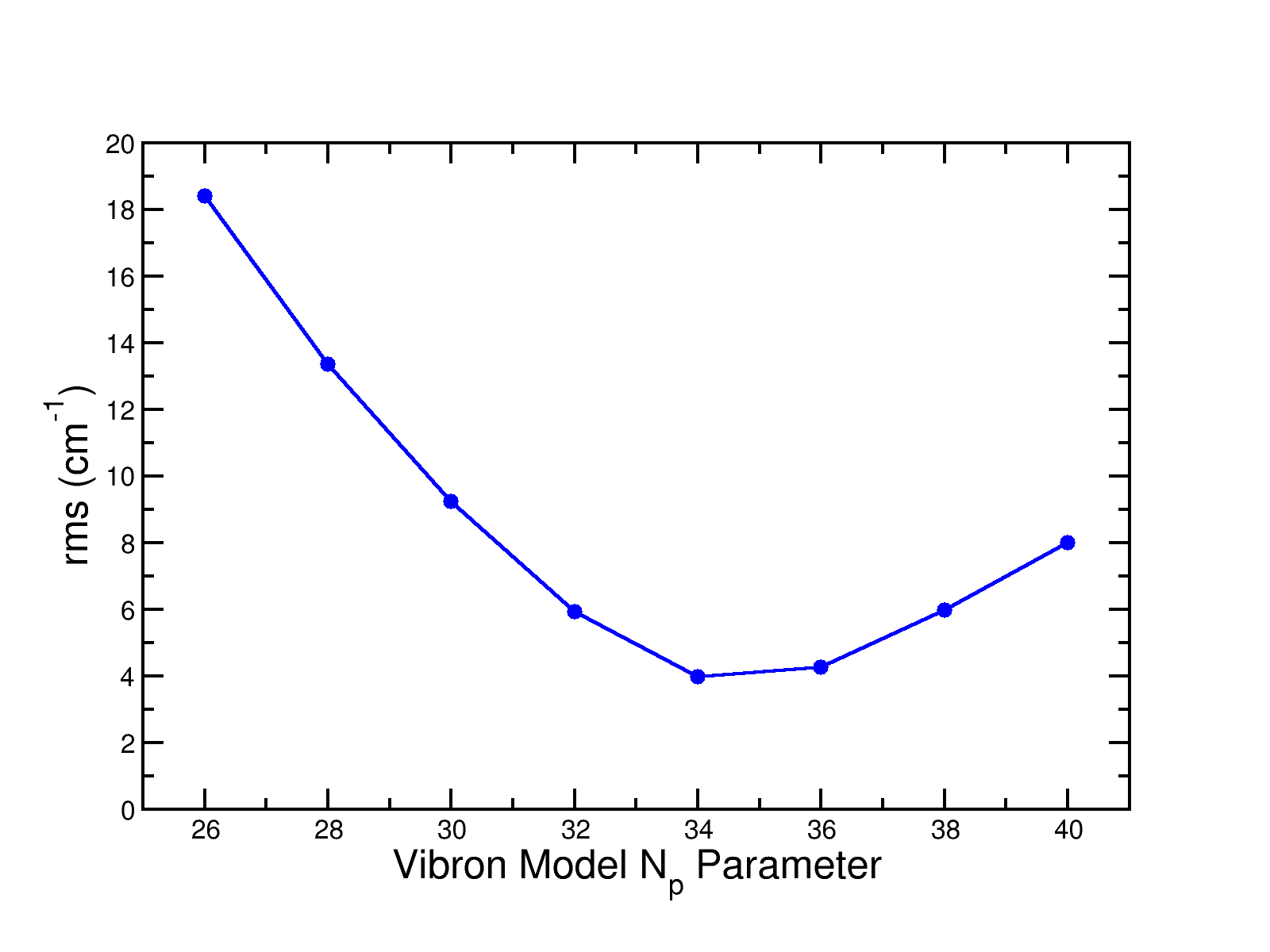}
\caption{(Color online) Root mean square deviation (\textit{rms}) for fits to free H$_2$ roto-vibrational experimental term energies under a threshold value of $10000\;\text{cm}^{-1}$ with Hamiltonian \eqref{hamvibronfreeH2} as a function of the number of vibrons parameter $N_p$.\label{figura_rms_Np}}
\end{figure}

The first step in the fitting procedure has been the assessment of the
parameter $N_p$. As experimental data for the endohedrally confined
species only involve $v=0,1$ H$_2$ vibrational
states, there is not enough information to estimate the $N_p$
parameter for the hydrogen molecule. This parameter is usually
fixed considering the ratio between first and second order parameters
in the Dunham expansion for the molecule under study
\cite{Iachello:94}. Therefore, we devised an alternative way to assess
this parameter, using the roto-vibrational spectroscopy of the free
H$_2$ molecule making use of free H$_2$ vibrational data and explored the $N_p$
dependence of the fit to the experimental energy terms beneath $10000$
cm$^{-1}$ with a $so(4)$ dynamical symmetry Hamiltonian
\begin{align}
\hat H_{so(4)}=&\beta\, \hat C_2[so(4)] + \gamma\, \hat C_2[so(3)]\label{hamvibronfreeH2} \\
&+  \gamma_2 \hat C_2[so(3)]^2+ \kappa\, \hat C_2[so(3)]\hat C_2[so(4)]~.\nonumber
\end{align}
The reason for setting an energy threshold is that the inclusion of
highly-excited energy levels, close to the molecular dissociation limit, implies the
necessity of including continuum effects and resonances that are out
of the scope of the vibron model, based on a $u(4)$ compact lie algebra
\cite{Iachello:81, Iachello:82, Kim:86}. The resulting root mean square (\textit{rms})
deviation for a fit to a total of 31 roto-vibrational experimental
term energies from Refs.\ \cite{Dabrowski:84,Stanke:08,Salumbides:11}
is depicted as a function of $N_p$ in Fig.\ \ref{figura_rms_Np}, where
it is clear that the best fit is obtained for $N_p = 34$ and the
resulting parameters can be found in Tab.~\ref{freeH2_fit}.

\begin{table}
  \caption{Parameters of Hamiltonian \eqref{hamvibronfreeH2} optimized to reproduce experimental term energies of the free H$_2$ molecule under $10000\;\text{cm}^{-1}$ with $N_p = 34$ and $rms$ of the fit. Parameters are given in cm$^{-1}$ units. The fit to 31 experimental energy levels has an $rms = 4.0\;\text{cm}^{-1}$.\label{freeH2_fit}}
    \begin{tabular}{|c|c|c|c|}
\hline
\hline
      $\beta$ & $\gamma$ &$\gamma_2$ &$\kappa$ \\
      $1041.54(6)$ & $32.80(7)$ &$-0.036215(7)$ &$0.72423(20)$ \\
\hline
\hline
    \end{tabular}
\end{table}

With the value of $N_p$ set to 34, we can return to the caged system. A Python code has been developed to calculate the eigenvalues and eigenstates of the total Hamiltonian  $\hat H_{endo}$ that encompasses the molecular roto-vibrational degrees of freedom \eqref{hup4}, the incarcerated center-of mass degrees of freedom \eqref{cageham}, and the coupling between them \eqref{hcoup}, and to compute the free parameter values that minimize the difference between calculated results and experimental line positions from Refs.\ \cite{Ge:11,Horsewill:12}. The code makes use of \textit{Sympy} \cite{sympy} and \texttt{LMFIT} \cite{lmfit} packages and is available under request.

In a preliminary set of calculations we have made several fits to the
full data set and to different subsets, obtaining a good overall
description. We have found that, leaving aside a constant energy shift,
a minimal Hamiltonian that complies with all symmetry requirements and
provides results that agree with experimental data, has seven
parameters: $\{\beta,\gamma,\kappa, a,b,c, \vartheta_{pq}\}$. The
first three from Eqs.\ \eqref{so4ham} and \eqref{corrham}, the second
three from Eq.\ \eqref{cageham}, plus the low order quadrupole
coupling in Eq.\ \eqref{hcoup}. The fits with this set are denoted as $F_0$. 

A finer fit, denoted as $F_1$, can be obtained with three more
parameters, up to a total of ten free parameters. The three added
parameters are $\gamma_2$ from Eq.\ \eqref{corrham}, and the
coupling parameters $\vartheta_{pqw}$ and $v_{pq}$ of Eq.\ \eqref{hcoup} associated with
operators $\left[\hat C_2[so_p(4)][\hat Q_p^{(2)}\times \hat
  Q_q^{(2)}]^{(0)}+[\hat Q_p^{(2)}\times \hat Q_q^{(2)}]^{(0)}\hat
  C_2[so_p(4)]\right]$ and $\hat C_1[u_q(3)]\hat C_2[so_p(4)]$,
respectively.

\begin{table}
\caption{Reassigned experimental transitions. Experimental states are given with the quantum numbers $vJN_qL\Lambda$. Line positions are given in cm$^{-1}$ units.\label{tabnewass}}
\begin{ruledtabular}
\begin{tabular}{cc|cc|cc}
\multicolumn{2}{c|}{Old ass.}&\multicolumn{2}{c|}{New ass.} & & \\
Initial & Final   &Initial & Final   & Exp. & Ref.\\
\hline
$00200$ & $01111$ &$00200$ & $01110$ & $-85.5$ &\cite{Horsewill:12} \\
$01221$ & $11311$ &$01334$ & $11443$ & $4294.8$ &\cite{Ge:11} \\
$00200$ & $10311$ &$01334$ & $11445$ & $4294.8$ &\cite{Ge:11} \\
$01334$ & $11444$ &$01221$ & $11311$ & $4300.0$ &\cite{Ge:11} \\
$01334$ & $11445$ &$01332$ & $12312$ & $4316.4$ &\cite{Ge:11} 
\end{tabular}
\end{ruledtabular}
\end{table}

\begin{figure}
\includegraphics[width=0.45\textwidth, bb= 0 0 500 900]{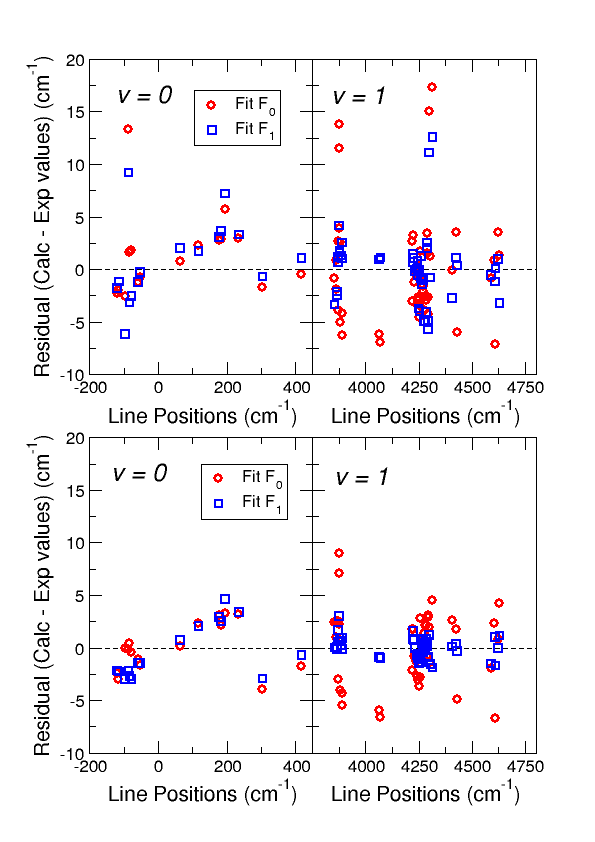}
\caption{Color online. Residuals of the $F_0$ and $F_1$ fits (see
  text) with the original assignments (upper panels) and including the
  changes suggested in Tab.~\ref{tabnewass} (lower  panels).\label{fig_residuals}}
\end{figure}

Preliminary calculations gave a satisfactory agreement with the
experimental line positions, though some levels had a residual value
much larger than expected from the overall fit agreement. This has
suggested us to consider a tentative reassignment of a set of five
transitions showing unusually large deviations, as indicated in
Tab.~\ref{tabnewass}. With this reassignment the
quality of the fit has largely improved. The convenience of this
reassignment in the framework of this model can be seen in Fig.\
\ref{fig_residuals} where the residuals for fits $F_0$ and $F_1$ are
plot as a function of the line position energy. The outcome for the
original level assignment is shown in the upper panels, while the
residuals with the new level assignment are depicted in the lower
panels. The achieved improvement in the fit is remarkable though a
deeper analysis is on the way to confirm these assignments and the
findings will be published in a forthcoming paper \cite{Perez:16tobe}. In
the following we refer to the set of experimental states with the five
mentioned reassignments.

The final $F_0$ and $F_1$ parameters, with  \textit{root mean
  square} $rms =
3.1\;\text{cm}^{-1}$ and  $1.7\;\text{cm}^{-1}$, respectively, are
given in Tab.~\ref{fit_pars}. The full list of residuals (experimental value minus
calculated value) for both fits, plotted in Fig.\ \ref{fig_residuals}
are given in Tab.~\ref{tab_residuals} together with the experimental
line positions and initial and final state assignments. 

\begin{table}
  \caption{$F_0$ (Minimal) and $F_1$ (Finer) fits parameter values. In both cases $N_p = 34$. Hamiltonian parameters and $rms$ are expressed in cm$^{-1}$ units.\label{fit_pars}}
  \begin{ruledtabular}
    \begin{tabular}{ccccc}
      $\hat H_{u_p(4)}$ &  $\beta$        &  $\gamma$   &  $\kappa$ &  $\gamma_2$ \\
      $F_0$       &  $-1083.23(18)$ & $58.09(17)$ &   $0.88(4)$ & --  \\
      $F_1$       &  $-1081.72(15)$ & $58.28(20)$ &   $0.810(25)$& $-0.032(15)$  \\
      \hline
      $\hat H_{u_q(3)}$ &  $a$            & $b$         &  $c$         &   \\
      $F_0$       & $178.3(8)$     & $9.6(3)$     &  $-3.26(15)$  &       \\
      $F_1$       & $179.0(4)$     & $8.46(17)$   &  $-3.18(8)$  &    \\
      \hline
      $\hat H_{Coupl}$ &  $\vartheta_{pq}$ & $\vartheta_{pqw}$ &      $v_{pq}$     &          \\
      $F_0$       &   $0.94(7)$& --      &  --           & \\
      $F_1$        &  $0.86(5)$& $-0.014(7)$& $-1.02(8)$  & \\
      \hline
      \hline
      $rms$ & $F_0$ & $3.1$ & $F_1$ &  $1.7$  \\
    \end{tabular}
  \end{ruledtabular}
\end{table}

\begin{figure*}
\includegraphics[width=0.1\textwidth, bb=100 0 290 229]{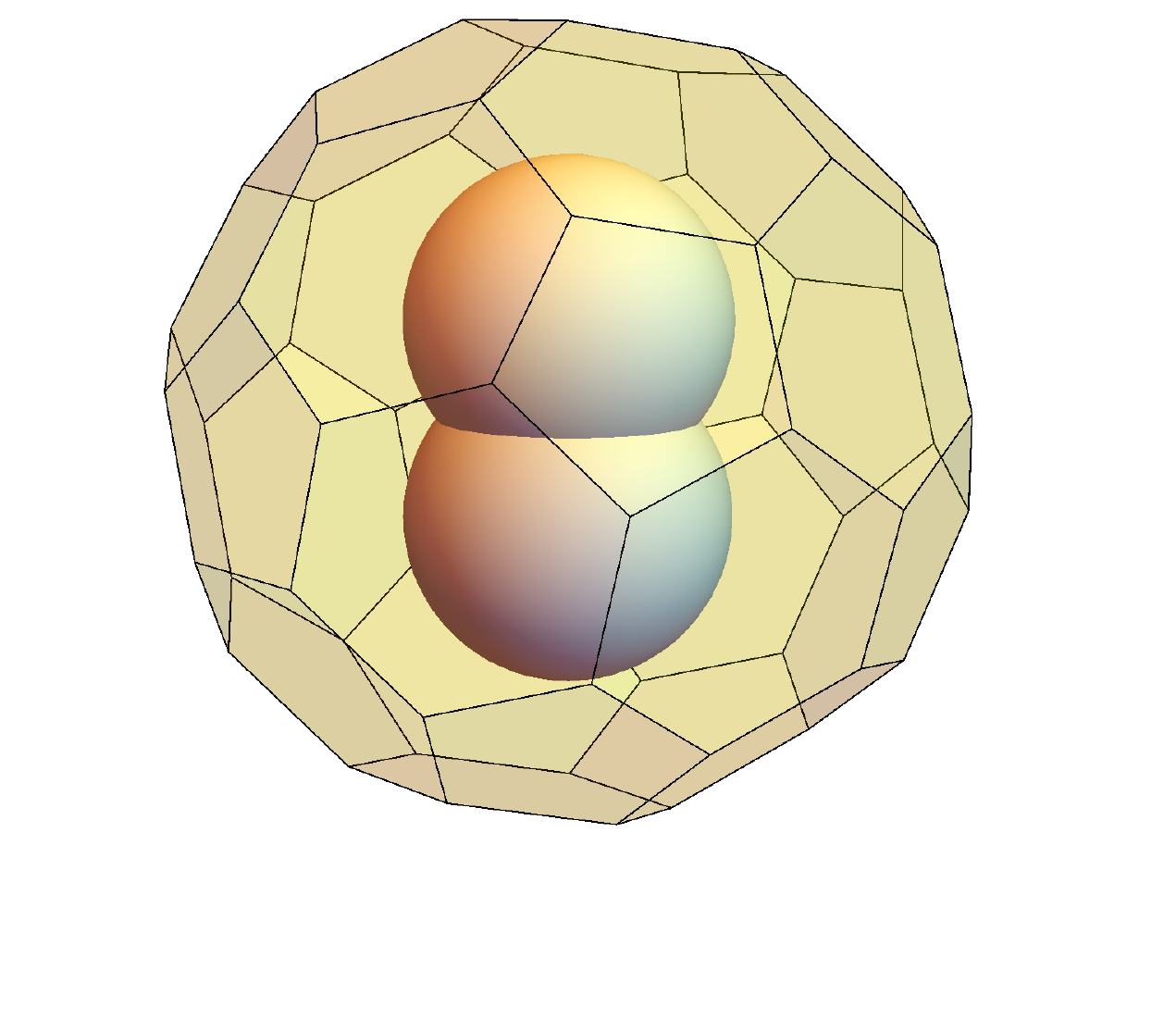}
\includegraphics[width=0.8\textwidth, bb=21 47 550 420]{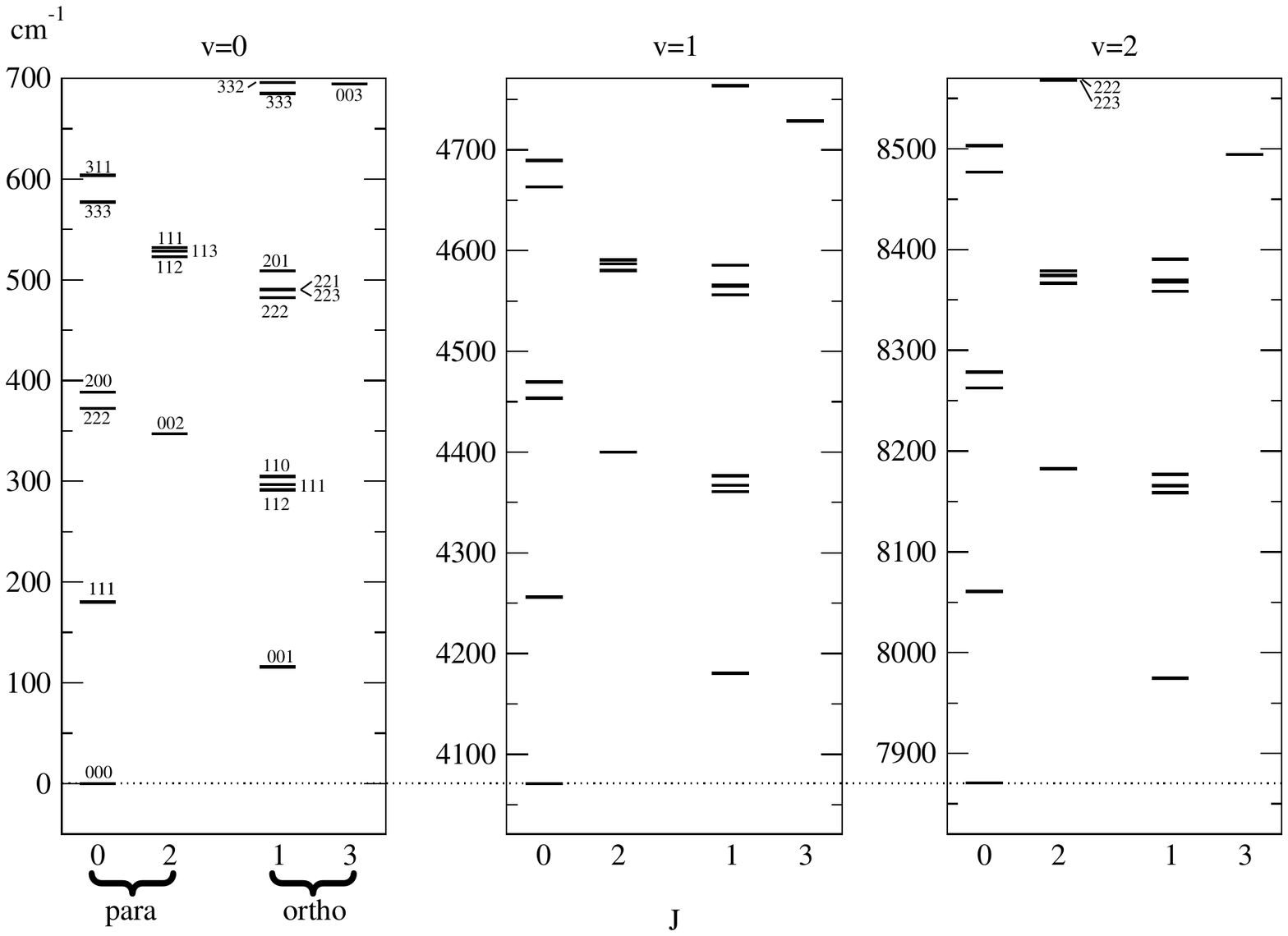}
\caption{Theoretical roto-vibrational spectrum of H$_2@$C$_{60}$. The
  three cuts show the energy levels in a 700 cm$^{-1}$ wide energy
  window just above the three lowest vibrational excitations $v=0,1$,
  and $2$. States are further divided into \textit{para} (left) and
  \textit{ortho} (right) states and are labeled by $J$ on the
  horizontal axis and $N_qL\Lambda$ on each state. These quantum
  numbers are repeated with the same order in each panel, except where
  noted.\label{figura}}
\end{figure*}

\begin{table*}
  \caption{Residuals for fits $F_0$ and $F_1$. Initial and final states are denoted by the quantum numbers of basis \eqref{basis}, $vJN_qL\Lambda$. Experimental states are extracted from Refs.\ \cite{Ge:11, Horsewill:12} and reassigned transitions (see Tab.\ I) are highlighted in red. Both experimental line positions and residual values are expressed in cm$^{-1}$.\label{tab_residuals}}
  \begin{ruledtabular}
    \begin{tabular}{c|c|c|c|c||c|c|c|c|c}
      Initial &      Final  &  Exp.  &  Calc.\ $F_0$ & Calc.\ $F_1$&Initial &      Final  &  Exp.  &  Calc.\ $F_0$ & Calc.\ $F_1$\\
\hline
0 1 0 0 1 & 0 0 0 0 0 &  -118.6& -2.42&  -2.16 & 0 1 0 0 1 & 1 1 1 1 1 &  4244.4& -2.76&  -0.96\\
0 1 1 1 1 & 0 0 1 1 1 &  -113.7& -3.00&  -2.27 & 0 2 1 1 2 & 1 2 2 2 3 &  4250.7& -0.65&  -1.12\\
0 0 2 0 0 & 0 1 1 1 2 &   -95.7& -0.06&  -3.00 & 0 1 0 0 1 & 1 1 1 1 2 &  4250.7& -3.14&  -1.48\\
\textcolor{red}{0 0 2 0 0} & \textcolor{red}{0 1 1 1 0} &   \textcolor{red}{-85.5}& \textcolor{red}{-0.17}&  \textcolor{red}{-2.19} & 0 0 0 0 0 & 1 0 1 1 1 &  4255.6& -3.649&  -1.45\\
0 0 2 2 2 & 0 1 1 1 1 &   -82.7&  0.44&  -2.67 & 0 1 1 1 2 & 1 1 2 2 2 &  4255.6& -0.84&  -0.82\\
0 0 2 2 2 & 0 1 1 1 2 &   -76.9& -0.44&  -2.97 & 0 1 0 0 1 & 1 1 1 1 0 &  4261.3& -2.85&  -1.43\\
0 2 0 0 2 & 0 1 1 1 1 &   -57.7& -1.09&  -1.46 & 0 2 1 1 1 & 1 2 2 2 0 &  4261.3&  2.77&   0.83\\
0 2 0 0 2 & 0 1 1 1 2 &   -51.6& -1.66&  -1.47 & 0 1 1 1 2 & 1 1 2 2 3 &  4267.1&  0.64&   0.43\\
0 1 0 0 1 & 0 0 1 1 1 &    65.2&  0.15&   0.73 & 0 1 1 1 1 & 1 1 2 2 1 &  4272.1& -1.05&  -0.46\\
0 0 0 0 0 & 0 1 0 0 1 &   118.5&  2.32&   2.06 & 0 0 1 1 1 & 1 0 2 2 2 &  4272.1&  0.25&   0.72\\
0 1 0 0 1 & 0 1 1 1 1 &   178.8&  3.05&   2.91 & 0 1 2 2 2 & 1 1 3 3 3 &  4277.1&  1.37&   0.46\\
0 1 0 0 1 & 0 1 1 1 2 &   184.5&  2.07&   2.51 & 0 1 2 2 3 & 1 1 3 3 4 &  4281.2&  2.12&   0.05\\
0 1 0 0 1 & 0 1 1 1 0 &   196.0&  3.26&   4.61 & 0 0 2 2 2 & 1 0 3 3 3 &  4286.5&  2.06&   0.81\\
0 1 0 0 1 & 0 2 0 0 2 &   235.5&  3.13&   3.38 & 0 1 1 1 2 & 1 1 2 0 1 &  4290.2&  0.62&   0.55\\
0 0 0 0 0 & 0 1 1 1 0 &   304.9& -4.02&  -2.92 & 0 0 1 1 1 & 1 0 2 0 0 &  4290.2& -0.79&   0.14\\
0 1 0 0 1 & 0 2 1 1 1 &   417.8& -1.84&  -0.71 & \textcolor{red}{0 1 3 3 4} & \textcolor{red}{1 1 4 4 3} &  \textcolor{red}{4294.8}&  \textcolor{red}{2.82}&  \textcolor{red}{-0.93}\\
0 1 2 0 1 & 1 1 1 1 2 &  3855.6&  2.37&   0.01 & \textcolor{red}{0 1 3 3 4} & \textcolor{red}{1 1 4 4 5} &  \textcolor{red}{4294.8}&  \textcolor{red}{3.10}&  \textcolor{red}{-0.83}\\
0 0 2 0 0 & 1 0 1 1 1 &  3866.0&  1.0&   0.08 & \textcolor{red}{0 1 2 2 1} & \textcolor{red}{1 1 3 1 1} &  \textcolor{red}{4300.0}&  \textcolor{red}{1.91}&   \textcolor{red}{1.19}\\
0 1 2 0 1 & 1 1 1 1 0 &  3866.0&  2.45&  -0.14 & 0 1 2 2 2 & 1 1 3 1 1 &  4306.7& -1.424&  -1.570\\
0 2 1 1 3 & 1 2 0 0 2 &  3872.2& -3.04&   0.59 & \textcolor{red}{0 1 3 3 2} & \textcolor{red}{1 2 3 1 2} &  \textcolor{red}{4316.4}&  \textcolor{red}{4.45}&  \textcolor{red}{-1.90}\\
0 1 2 2 1 & 1 1 1 1 1 &  3872.2&  2.53&   1.68 & 0 1 2 0 1 & 1 3 1 1 2 &  4407.4&  2.63&   0.11\\
0 1 3 3 3 & 1 1 2 2 2 &  3876.0&  8.92&   3.02 & 0 1 2 2 3 & 1 3 1 1 4 &  4426.8&  1.77&   0.39\\
0 0 3 3 3 & 1 0 2 2 2 &  3878.6&  7.04&   0.67 & 0 1 1 1 2 & 1 3 0 0 3 &  4431.9& -4.94&  -0.30\\
0 1 2 2 3 & 1 1 1 1 2 &  3878.6&  2.25&   0.96 & 0 0 0 0 0 & 1 2 1 1 1 &  4592.0& -2.02&  -1.53\\
0 0 2 2 2 & 1 0 1 1 1 &  3884.9&  0.72&   0.20 & 0 0 1 1 1 & 1 2 2 2 1 &  4608.9&  2.25&   1.02\\
0 1 1 1 2 & 1 1 0 0 1 &  3884.9& -4.08&   0.62 & 0 1 0 0 1 & 1 3 0 0 3 &  4612.5& -6.77&  -1.69\\
0 1 1 1 1 & 1 1 0 0 1 &  3891.3& -4.36&   0.92 & 0 0 1 1 1 & 1 2 2 0 2 &  4624.3&  0.76&  -0.06\\
0 0 1 1 1 & 1 0 0 0 0 &  3891.3& -5.49&  -0.16  & 0 0 2 2 2 & 1 2 3 3 1 &  4630.0&  4.24&   1.14\\
0 1 0 0 1 & 1 1 0 0 1 &  4065.4& -6.01&  -0.86 & 0 1 0 0 1 & 1 3 1 1 2 &  4802.6& -2.77&  -1.28\\
0 0 0 0 0 & 1 0 0 0 0 &  4071.4& -6.62&  -0.96 & 0 1 1 1 2 & 1 3 2 2 2 &  4814.8&  0.13&  -0.17\\
0 3 0 0 3 & 1 3 1 1 4 &  4223.3&  1.70&   0.76 & 0 1 1 1 1 & 1 3 2 2 2 &  4821.6&  0.26&   0.53\\
0 0 1 1 1 & 1 2 0 0 2 &  4223.3& -2.20&   1.57 & 0 1 1 1 1 & 1 3 2 2 1 &  4829.7&  1.29&   1.40\\
0 3 0 0 3 & 1 3 1 1 2 &  4226.2&  1.74&   0.73 & 0 1 1 1 2 & 1 3 2 0 3 &  4836.2&  1.68&   1.74\\
0 2 0 0 2 & 1 2 1 1 2 &  4233.1& -0.82&  -0.04 & 0 1 2 2 2 & 1 3 3 3 1 &  4846.4&  1.83&   0.35\\
0 2 0 0 2 & 1 2 1 1 3 &  4239.8& -1.34&  -0.72 & 0 1 2 2 1 & 1 3 3 1 2 &  4864.5& -0.96&  -2.24\\
0 2 0 0 2 & 1 2 1 1 1 &  4244.4& -1.07&  -0.56 &           &           &        &       &        \\
\end{tabular}
\end{ruledtabular}
\end{table*}

The quality and robustness of our calculations allows us to estimate the energies of
levels not yet accessed experimentally. We have included in Fig.\
\ref{figura} the calculated $v = 0,1$, and $2$ levels, the latter not
yet measured. One can notice that, with growing $v$, the higher the
$J$ the bigger the negative energy shift of corresponding states. An extensive
table with all calculated levels for $v=0,1,2$ vibrational quanta with
$N_q\le 4$ and $\Lambda \le 5$ can be found in the Supplemental
Material section. In addition to the term energy, expressed in
cm$^{-1}$ units, we also indicate in the table the probability of the largest component (squared
coefficient) of the corresponding eigenstate expressed in basis
\eqref{basis}.



\section{Summary and Conclusions}
In summary, we have introduced a $u(4)\oplus u(3)$ algebraic scheme
that combines the vibron model description of a diatomic molecule
roto-vibrational structure with an algebraic description of the motion
of the molecule center-of-mass inside an isotropic cage. This model is
mathematically rich and has a large number of possible terms that can
be attributed to different physical mechanisms. We have presented here
a discussion of a few selected physical mechanisms that, in spite of
the model's simplicity, give insight into the spectroscopic properties
of diatomic endohedrally confined molecules. We have then applied the
symmetry-guided scheme to a database of experimental lines, finding a
very good overall agreement to the experimental line positions and
finding that the fits improve considerably upon reassigning the
quantum numbers of a small subset of energy levels.

The next step is the inclusion of transition intensities in the model
and the enrichment of the approach, that could take place in one of
two possible venues, either by defining an embedding $u_{pq}(7)
\supset u_{p}(4) \oplus u_{q}(3)$ dynamical algebra or by a dynamical
algebra $u_{p}(4)\oplus u_{q}(4)$ with results that will be published
soon \cite{Perez:16tobe}. Another venue for future research is the
inclusion in the algebraic model of the cage icosahedral symmetry
effect on the spectrum, that has recently been
investigated \cite{Poirier:15,Mamone:16}.

\begin{acknowledgments}
LF acknowledges financial support within the PRAT~2015 project {\it IN:Theory}, Univ. of Padova (project code: CPDA154713). FPB was funded by MINECO grant FIS2014-53448-C2-2-P.  We thank Jos\'e M.\ Arias, Alejandro Frank, Francesco Iachello, and Renato Lemus for useful discussions and valuable suggestions.   
\end{acknowledgments}

\bibliographystyle{apsrev4-1}
%

\end{document}